\documentclass[aps,prb,twocolumn,superscriptaddress,amsmath,amssymb,showpacs]{revtex4}
\usepackage{psfrag}
\usepackage{bm}
\usepackage{graphicx}
\usepackage{subfigure}
\usepackage{dcolumn}
\newcommand{\sgn}{\text{sgn}}

\begin{document}

\title{Phase diagram for quantum Hall states in graphene}

\author{Jianhui Wang}
\affiliation{Department of Physics, Indiana University,
Bloomington, Indiana 47405, USA}
\author{A. Iyengar}
\affiliation{Department of Physics, Indiana University,
Bloomington, Indiana 47405, USA}
\author{H. A. Fertig}
\affiliation{Department of Physics, Indiana University,
Bloomington, Indiana 47405, USA}
\affiliation{Department of Physics, Technion, Haifa 32000, Israel}
\author{L. Brey}
\affiliation{Instituto de Ciencia de Materiales de Madrid (CSIC), Catoblanco, 28049 Madrid, Spain}

\date{\today}

\begin{abstract}
We investigate integer and half-integer filling states (uniform
and unidimensional stripe states respectively) for graphene using
the Hartree-Fock approximation. For fixed filling factor, the
ratio between the scales of the Coulomb interaction and Landau
level spacing $g=(e^2/\epsilon \ell)/(\hbar v_F/\ell)$, with
$\ell$ the magnetic length, is a field-independent constant.
However, when $B$ decreases, the number of filled negative Landau
levels increases, which surprisingly turns out to decrease the
amount of Landau level mixing. The resulting states at fixed
filling factor $\nu$ (for $\nu$ not too big) have very little
Landau level mixing even at arbitrarily weak magnetic fields. Thus
in the density-field phase diagram, many different phases may persist down to
the origin, in contrast to the more standard two dimensional
electron gas, in which the origin is surrounded by Wigner crystal states.
We demonstrate that the stripe amplitudes scale
roughly as $B$, so that the density waves ``evaporate''
continuously as $B\rightarrow 0$. Tight-binding calculations give
the same scaling for stripe amplitude and demonstrate that the
effect is not an artifact of the cutoff procedure used in the
continuum calculations.
\end{abstract}


\maketitle

\section{\label{sec:into}INTRODUCTION}
\begin{figure*}
\begin{center}
\subfigure[2DEG]{\label{fig:2DEG}\includegraphics[scale=0.725]{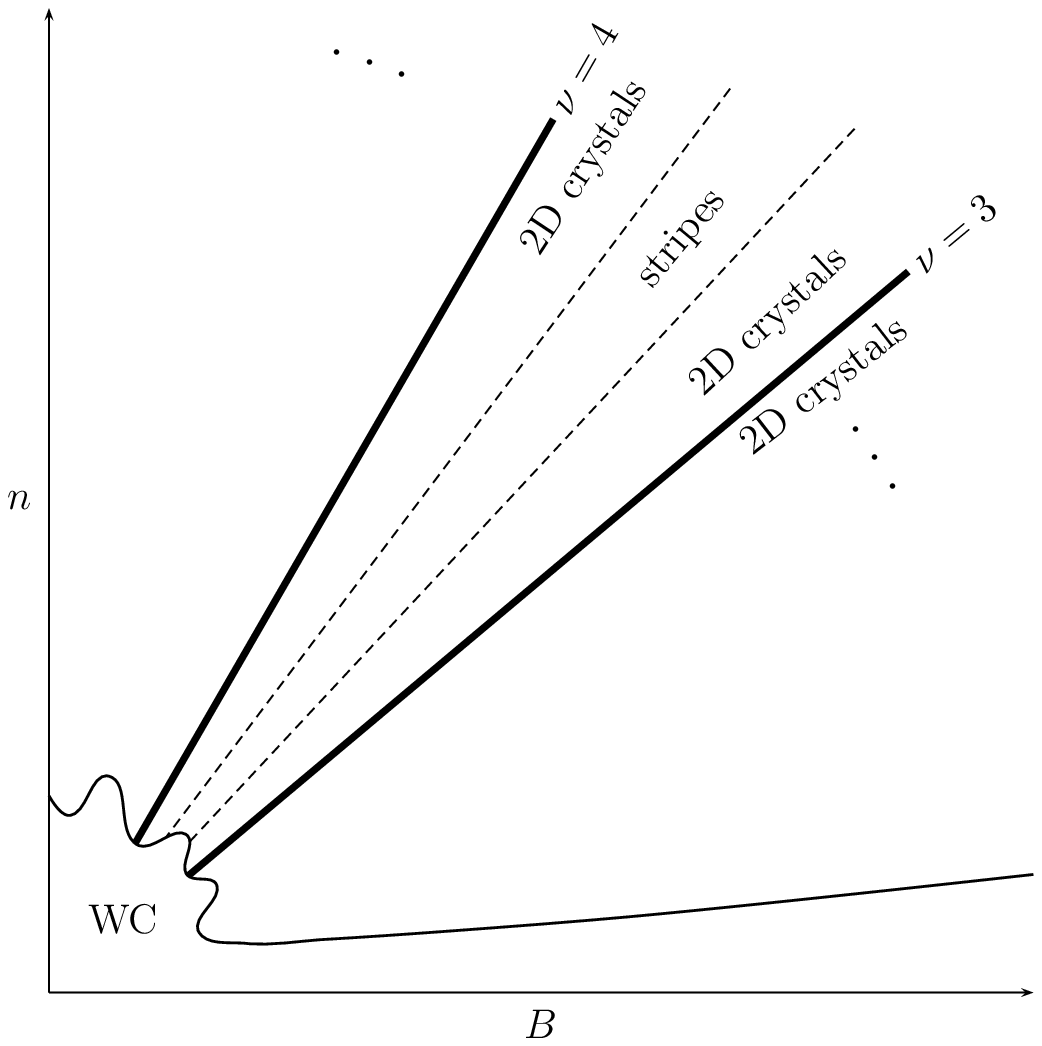}}
\subfigure[graphene]{\label{fig:graphene}\includegraphics[scale=0.725]{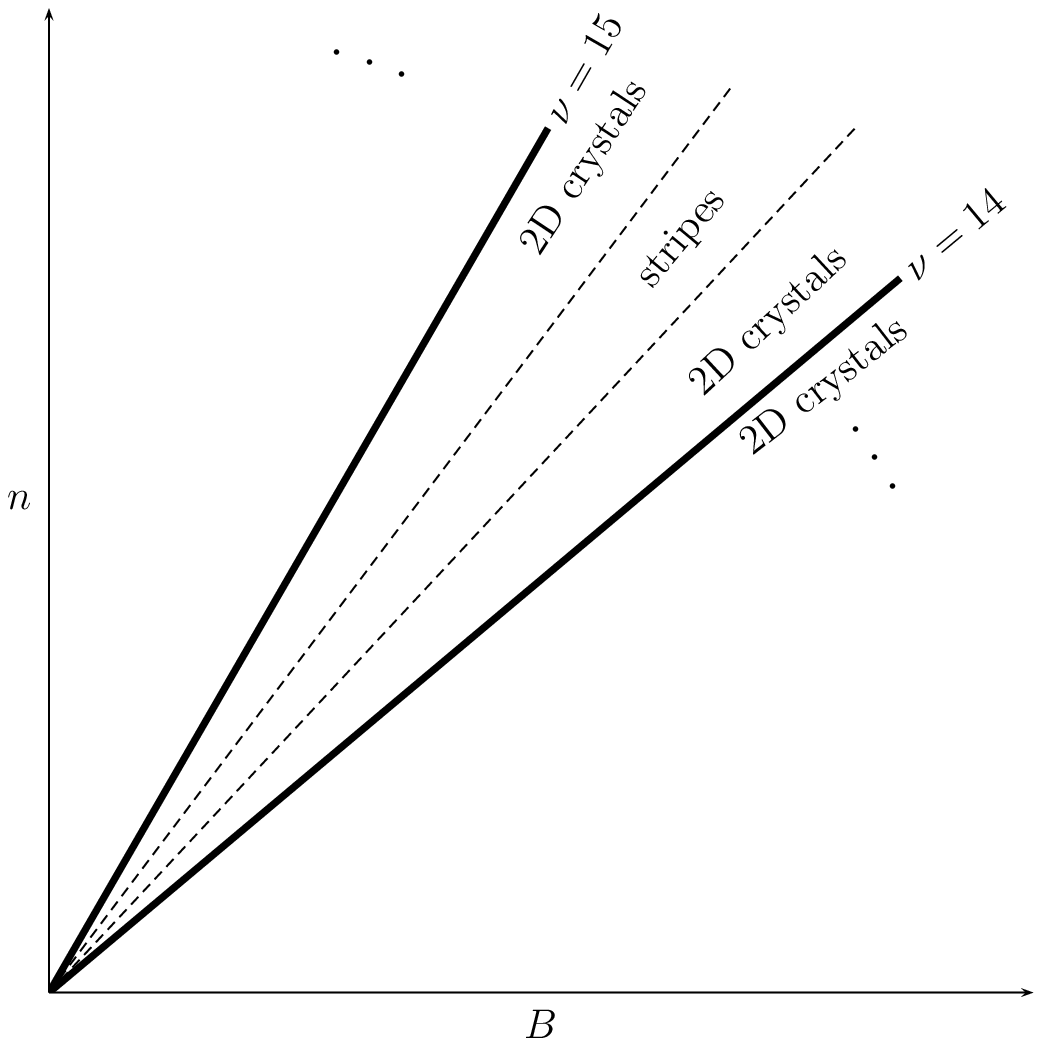}}
\end{center}
\caption{Schematic phase diagrams for conventional  2DEG and
graphene in the integer quantum Hall regime. Here $n$ is the electron
density. ``2D crystals'' referred to in diagrams
include bubble, quasiparticle and quasihole states, whose lattice
constants are determined by the filling factor and magnetic length.  These differ
qualitatively from the Wigner crystal state where the lattice
constant is set by the electron density.  Fractional quantum
Hall states, expected to appear at low filling factors in
both diagrams, are not shown.
The major difference
between the two phase diagrams is that for the conventional 2DEG,
the origin is
completely surrounded by the Wigner crystal state, while in the graphene
case, many different phases persist down to the origin.}
\label{fig:diagrams}
\end{figure*}

Graphene, a two-dimensional honeycomb lattice of carbon atoms, has
attracted intense attention in the past few years.\cite{Neto2007}
Its properties bear some similarities with, and some striking
differences from, conventional 2D electron gas (2DEG) systems
found in semiconductor heterostructures. It is well-known that the
latter has a rich phase diagram in the quantum Hall regime. When
$r_s$, the average inter-electron distance measured in units of
Bohr radius, is not very big, there are integer and fractional
quantum Hall liquid states, as well as charge density waves (CDWs)
of various forms, including Wigner crystals of quasi-electrons,
bubbles and stripes at fillings around these liquid states, and
analogous particle-hole conjugates of these states
\cite{Fradkin1999,Yi2000,Cote2004}. In high magnetic fields, the
particular state is essentially determined by the filling factor
$\nu$, defined as the ratio of the electron density to the density
of magnetic flux quanta penetrating the plane. When $r_s$ is
increased, these quantum Hall phases undergo transitions to Wigner
crystal (WC) states with a single electron per unit cell. (For
very small $\nu$, there may also be Wigner crystals of composite
fermions \cite{Yi1998,Mandal2003}.) If the phase diagram is
plotted in the $n$ (density) - $B$ (magnetic field) plane, away
from the origin, there is a fan of quantum Hall phases, but the
origin is expected to be completely surrounded by Wigner crystal
states \cite{Csathy2005} [see Fig.~\ref{fig:2DEG}].

The integer quantized Hall effect has been observed in
graphene \cite{novoselov05,zhang05,zhang06,abanin07},
and, except for a well-understood shift in the precise
values of the plateaus \cite{Ando-review, gusynin05}, the
Hall conductance appears rather similar to that
found in the conventional 2DEG.  Nevertheless,
the behavior of clean and cold graphene in the low doping
limit is likely to be different than that of the conventional 2DEG. Unlike
the latter, non-interacting electrons in graphene to a good approximation obey
a massless Dirac equation \cite{Ando-review,Gusynin-review,Castro-review},
with two inequivalent Dirac points in two different valleys (denoted
as ${\bf K}$ and ${\bf K^{\prime}}$) in the
Brillouin zone.  When the system is undoped the
Fermi energy passes directly through these Dirac points.
With interactions, continuum \cite{DahalHP2006} and
tight-binding \cite{Dahal2007} studies of the this system
in mean-field theory indicate that the system remains
in a liquid state in zero magnetic field even at arbitrarily low doping.
On
the other hand, Hartree-Fock
calculations \cite{ZhangCH2007} and exact
diagonalization studies \cite{WangHao2007}
suggest that CDWs are possible in a large magnetic field -- where
states are restricted to a single or two \cite{ZhangCH2008} Landau levels (LLs) --
and that the phase diagram is similar to
that of the conventional 2DEG.
In this paper we address the question
of how the system passes
from these strong-field states
into the liquid state as the field and density are lowered to
small values.

For the conventional 2DEG, the quantum Hall states give
way to the WC in the low-field, low density limit due to
Landau level mixing (LLM). This allows the
electrons to form wavepackets that are more localized
than is possible within a single Landau level, thereby
lowering the interaction energy \cite{Zhu1993}.
The degree of LLM is determined by a coupling constant $g$,
the ratio of the typical Coulomb
interaction energy $E_C$ to the scale of the LL separation. For both
graphene and the conventional 2DEG, $E_C$ is given by $e^2/(\epsilon
\ell)$, where $\ell=\sqrt{\hbar c/e B}$ is the magnetic length.
However, the LL separations are  different in the two cases. In the
conventional 2DEG, it is given by $\hbar \omega_c=\hbar e B/m c$,
so that $g \propto 1/\sqrt{B}$ and in the large $B$
limit where $g$ is small, LLM is negligible.
In graphene, the LLs are
not equally spaced \cite{Ando-review}, so we instead
characterize it by the gap between the $n=0$ and $n=1$ LLs divided by $\sqrt{2}$,
$\hbar v_F/\ell$,
where $v_F$ is the Fermi velocity. Then
$g=(e^2/\epsilon)/(\hbar v_F)$ is a field-independent constant \cite{DahalHP2006},
typically
estimated to be of order 1 or smaller.
Nevertheless, even though $g$ is field-independent, the degree of LLM
can change with $B$ even for fixed filling factor, and we shall
see below that it in fact does, albeit by a small amount.
This is because in addition to positive energy levels, the Dirac equation
in a magnetic field supports negative energy Landau level states, as well
as a zero energy LL, for each spin, provided the Zeeman
energy is neglected.  Moreover, the low energy theory of
graphene involves two such Dirac points
(${\bf K}$ and ${\bf K^{\prime}}$ valleys), so there are
two copies of these energy levels in the spectrum.  When undoped, 
all the negative energy states are filled, as well as one of the two
zero-energy states \cite{Ando-review}.  Added electrons interact with
the electrons in these filled levels, which changes the {\it effective}
energy of the higher LLs.
Because the Landau level structure of these filled levels
varies with field, the splitting between the $n=0$ and $n=1$ energy levels
does not precisely the follow the $\sqrt{B}$ behavior discussed above.

In a continuum description, the filling of the negative levels is
characterized by a (negative) integer $n_c$, which denotes the lowest
LL which must be filled to accommodate one electron per atom, the
density of mobile electrons of undoped graphene \cite{com,Iyengar2007}.
An extra field dependence thus enters the problem through $n_c$,
and is given by
\begin{equation}
-n_c
=\frac{2 S/(\sqrt{3} a^2/2)}{4 S /2 \pi \ell^2}\propto
\frac{1}{B}, \label{cutoff}
\end{equation}
where $S$ is the area of the sample, $a=0.246\,\text{nm}$  is the
lattice constant of the triangular (Bravais) lattice and
the factor of 4 in the denominator comes from
the spin and valley degeneracies.
The interaction of electrons with those in the negative levels
is most easily described in the
Hartree-Fock (HF) approximation, where it appears as a
contribution to the exchange self-energy.
For uniform liquid states, we find that
the Coulomb energy decreases faster with decreasing B than the difference
in effective energy
between the highest occupied level and the lowest
unoccupied level, so that the ratio between kinetic and potential energy
actually {\it increases} with decreasing
$B$, because $|n_c|$ increases.  We will demonstrate a similar
effect for stripe states,
and believe it should be
ubiquitous for charge-ordered and liquid quantum Hall states.

Because this effect is a result of filling $|n_c|$ negative LLs,
it is a concern that it may be an artifact of the cutoff procedure
used in our HF calculations.  To check this, we performed an
analogous calculation for interacting electrons in a tight-binding
model, where no artificial cutoff needs to be introduced.  We
obtain results from this model that are
very similar to those of the
continuum model.

The consequence of this is that, for states where LLM is
small at large values of $B$, we expect it remain small,
and even {\it decrease}, with decreasing $B$.  While it is
not immediately obvious that with $g \sim 1$ one should find
weak LLM in these quantum Hall states, this does appear to
be the case for WC and bubble states \cite{ZhangCH2008},
and as we demonstrate below, for stripe and uniform
liquid states.
The surprising result is that,
within the Hartree-Fock approach, one expects
these states to persist to
arbitrarily small field. Thus,
many different states persist down to the origin of the phase
diagram in $n$ - $B$ plane [see Fig.~\ref{fig:graphene}].
Because these states follow trajectories of fixed $\nu$ in
the plane, the density of electrons participating these
CDW states decreases with $B$, such that their
amplitude scales roughly as $B$,
and the wavelength as $1/\sqrt{B}$.
The stripe states, and by analogy other CDW states,
disappear continuously as $B\rightarrow 0$, eventually
becoming indistinguishable from a uniform liquid state
in the low field limit.  Nevertheless, in principle,
for a clean, undeformed graphene system, this implies
that in principle many states emanate from the $B=n=0$ point in the phase diagram.

We note that this behavior is very specific to the $1/r$ form
for Coulomb interactions that is natural in this system.  For shorter
range interactions, where a length scale other than the magnetic
length is involved in the interaction range, the effective value
of $g$ will increase with decreasing field as in the standard
2DEG, at low densities and fields LLM should destabilize the high field
states, and a WC state should result.  Such a situation could arise
if a metallic gate is sufficiently close to the graphene plane
to effectively screen the long-range component of the Coulomb interaction.

More generally, the behavior discussed here may be understood as being
a consequence of the marginal nature of the $1/r$ Coulomb interaction
in undoped graphene.  As has been shown by us elsewhere \cite{Iyengar2007},
the energy difference between Landau levels near the Fermi energy
is increased by the filled Fermi sea, by an amount proportional
to $\log|n_c|$.  This logarithmic increase of the LL spacing
with increasing $|n_c|$
can be reinterpreted as the Fermi velocity being renormalized
upwards as the high-energy cutoff of the theory is increased
\cite{gonzalez94}.  That the LL spacing
{\it increases} slightly with increasing $|n_c|$ as
the doping is decreased is consistent
with interactions being marginally irrelevant in this system \cite{gonzalez94}.
Had it decreased instead,  the
interactions would be marginally relevant,
one would expect to find a WC
state near the origin of the $n-B$ phase diagram.

This paper is organized as follows.  In
Section~\ref{sec:continuum}, we describe the continuum limit
Hamiltonian, and the Hartree-Fock approximation used to
study quantum Hall states in the presence of LLM. In
Section~\ref{sec:result}, we discuss the results of these continuum
calculations. In Section~\ref{sec:hubbard}, we introduce a
tight-binding model with Hubbard interactions, and demonstrate that the
suppression of LLM found in the continuum calculations is not
an artifact of our cutoff procedure.  Finally, we conclude with a
summary in
Section~\ref{sec:sum}.

\section{\label{sec:continuum} Hartree-Fock for Continuum Model}
In standard 2DEG's, it is known that the Hartree-Fock approximation
is quite reliable for electronic states in high Landau levels \cite{koulakov06,moessner06}.
The situation should be similar for graphene, particularly if
one can show that LLM is small for states in high LLs, as we
will indeed find self-consistently below.  We thus adopt the
Hartree-Fock approximation for the states we study.

More specifically,
our Hartree-Fock approach to the Dirac equation description of
uniform and stripe quantum Hall phases in graphene is adapted from
a procedure developed for electrons in a standard 2DEG
\cite{CoteR1991}; in what follows we briefly outline the method,
and highlight the (largely technical) differences. The Hamiltonian
for the system in a magnetic field is
\begin{equation}
\hat{H}=\sum_{1}\varepsilon_1 \hat{c}_1^\dagger \hat{c}_1+\frac{1}{2}\sum_{1234}v_{1234}\hat{c}_1^\dagger\hat{c}_2^\dagger\hat{c}_3\hat{c}_4,
\end{equation}
where the numbers denote composite indices for the different
quantum numbers specifying the states [e.g.,  $1\equiv
(n_1,\,X_1,\,s_1,\,t_1) =$ (LL index, guiding center coordinate,
spin, pseudospin (valley) index)],
\begin{equation}
\label{bareEnergy}
\varepsilon _1=\varepsilon _{n_1}^{s_1}=\sgn(n_1)\frac{\hbar v_F}{\ell} \sqrt{2 |n_1|}-s_1 g^\star \mu_B B
\end{equation}
is the LL spectrum plus the Zeeman energy, and
\begin{equation}
v_{1234}=\frac{1}{4} W_{1234}\delta_{s_1s_4}\delta_{t_1t_4}\delta_{s_2s_3}\delta_{t_2t_3}\prod_{i=1}^4 (\sqrt{2})^{\delta_{n_i,0}}
\end{equation}
are matrix elements for the Coulomb interaction. $W_{1234}$ is
related to standard matrix elements \cite{CoteR1991}
\begin{eqnarray}
\label{eq:Vtilde}
\tilde{V}_{n_1,n_2,n_3,n_4}&=&\frac{1}{S}\sum_{\bm{q}} V(\bm{q}) \left< n_1,X_1|e^{i \bm{q} \cdot \bm{r}}|n_4,X_4\right>\\ \nonumber
&&\times\left<n_2,X_2|e^{-i \bm{q} \cdot \bm{r}}|n_3,X_3\right>,
\end{eqnarray}
with $V(\bm{q})=2\pi e^2/q$ and
\begin{eqnarray*}
\lefteqn{\left< n_1,X_1|e^{i \bm{q} \cdot \bm{r}}|n_4,X_4\right>}\\
&=&\exp[i\frac{1}{2}q_x(X_1+X_4)]F_{n_1,n_4}(\bm{q})\delta_{X_1,X_4+q_y\ell^2},
\end{eqnarray*}
where
\begin{eqnarray*}
F_{n_1,n_4}(\bm{q})&=&\left(\frac{n_4!}{n_1!}\right)^{1/2}\left(\frac{(-q_y+iq_x)\ell}{\sqrt{2}}\right)^{n_1-n_4}\\
&&\times \exp\left(\frac{-q^2\ell^2}{4}\right)L^{n_1-n_4}_{n_4}\left(\frac{q^2\ell^2}{2}\right)
\end{eqnarray*}
for $n_4\leq n_1$, where $L^{\alpha}_n(x)$ is the generalized Laguerre polynomial. Note that $F_{n_4,n_1}(\bm{q})=[F_{n_1,n_4}(-\bm{q})]^{\ast}$.

In terms of $\tilde{V}$, $W$ takes the form
\smallskip
\begin{eqnarray}
\label{eq:W}
W_{1234}&=&(-i)^{|n_1|+|n_2|}i^{|n_3|+|n_4|} [\tilde{V}_{|n_1|,|n_2|,|n_3|,|n_4|}\\ \nonumber
&&+\sgn(n_1n_4) \tilde{V}_{|n_1|-1,|n_2|,|n_3|,|n_4|-1}\\ \nonumber
 & &+\sgn(n_2n_3) \tilde{V}_{|n_1|,|n_2|-1,|n_3|-1,|n_4|}\\ \nonumber
&&+\sgn(n_1n_2n_3n_4)\tilde{V}_{|n_1|-1,|n_2|-1,|n_3|-1,|n_4|-1}].
\end{eqnarray}
Note that the guiding center coordinates ($X$) have been
suppressed in the subscripts in Eqs.~(\ref{eq:Vtilde}) and (\ref{eq:W}). The density matrix operators are defined as
\begin{eqnarray}
\hat{\rho}^{nst}_{n's't'}(\bm{q})&\equiv & \frac{2\pi \ell^2}{S} \sum_ X \exp (-iq_x X-\frac{1}{2} i q_x q_y \ell^2)\\ \nonumber
&& \times \hat{c}^\dagger_{nXst} \hat{c}_{n'\,X+q_y \ell^2\,s't'}.
\end{eqnarray}
This relation may be inverted to obtain the expectation value of
an arbitrary single-particle operator in terms of density operator expectation values,
\begin{eqnarray}
\left<\hat{c}^\dagger_{nXst}\hat{c}_{n'X's't'}\right>&=&\sum_{\bm{p}}\left<\hat{\rho}^{nst}_{n's't'}(\bm{p})\right>\\ \nonumber
&&\times \exp[\frac{1}{2}i p_x (X+X')]\delta_{X,X'-p_y \ell^2}.
\end{eqnarray}
For states with discrete translational symmetry,
the sum over ${\bf p}$ is restricted to reciprocal lattice vectors $\left\lbrace {\bf G} \right\rbrace$.
The interaction part of the HF Hamiltonian \(\hat{H}_{HF}\) may now be written as
\begin{widetext}
\begin{equation}
\hat{H}_{e-e}=\frac{S}{2\pi\ell^2}\sum_{n_2,n_3}\sum_{\bm{G}}\sum_{s_2,t_2}[U_H(n_2,n_3;\bm{G})\hat{\rho}^{n_2s_2t_2}_{n_3s_2t_2}(\bm{G})-\sum_{s_1,t_1}U_X(s_1,s_2,t_1,t_2,n_2,n_3;\bm{G})\hat{\rho}^{n_2s_2t_2}_{n_3s_1t_1}(\bm{G})],
\end{equation}
where
\begin{eqnarray}
U_H(n_2,n_3;\bm{G})&\equiv& \frac{e^2}{4\ell}\sum_{n_1,n_4}\sum_{s_1,t_1}H_g(n_1,n_2,n_3,n_4;\bm{G})\left<\hat{\rho}^{n_1s_1t_1}_{n_4s_1t_1}(-\bm{G})\right>,\\
U_X(s_1,s_2,t_1,t_2,n_2,n_3;\bm{G})&\equiv& \frac{e^2}{4\ell}\sum_{n_1,n_4}X_g(n_1,n_2,n_3,n_4;\bm{G})\left<\hat{\rho}^{n_1s_1t_1}_{n_4s_2t_2}(-\bm{G})\right>,
\end{eqnarray}
with
\begin{eqnarray*}
\lefteqn{H_g(n_1,n_2,n_3,n_4;\bm{G})}\\ \nonumber
&\equiv &(-i)^{|n_1|+|n_2|} i^{|n_3|+|n_4|} \prod_{i=1}^{4}(\sqrt{2})^{\delta_{n_i,0}} [H(|n_1|,|n_4|,|n_2|,|n_3|;\bm{G})+\sgn(n_1n_4) H(|n_1|-1,|n_4|-1,|n_2|,|n_3|;\bm{G})\\ \nonumber
& &+\sgn(n_2n_3) H(|n_1|,|n_4|,|n_2|-1,|n_3|-1;\bm{G})+\sgn(n_1n_2n_3n_4)H(|n_1|-1,|n_4|-1,|n_2|-1,|n_3|-1;\bm{G})],
\end{eqnarray*}
\begin{eqnarray*}
\lefteqn{X_g(n_1,n_2,n_3,n_4;\bm{G})}\\
&\equiv &(-i)^{|n_1|+|n_2|} i^{|n_3|+|n_4|} \prod_{i=1}^{4}(\sqrt{2})^{\delta_{n_i,0}} [X(|n_1|,|n_3|,|n_2|,|n_4|;\bm{G})+\sgn(n_1n_3)X(|n_1|-1,|n_3|-1,|n_2|,|n_4|;\bm{G})\\ \nonumber
& &+\sgn(n_2n_4)X(|n_1|,|n_3|,|n_2|-1,|n_4|-1;\bm{G})+\sgn(n_1n_2n_3n_4)X(|n_1|-1,|n_3|-1,|n_2|-1,|n_4|-1;\bm{G})],
\end{eqnarray*}
where
\begin{eqnarray}
H(n_1,n_2,n_3,n_4;\bm{G})=\frac{1}{2\pi e^2\ell}V(\bm{G})F_{n_1,n_2}(\bm{G})F_{n_3,n_4}(-\bm{G}),\\
X(n_1,n_2,n_3,n_4;\bm{G})=\frac{\ell}{e^2 S}\sum_{\bm{q}}V(\bm{q})F_{n_1,n_2}(\bm{q})F_{n_3,n_4}(-\bm{q})\exp(-i\bm{q}\times \bm{G}\ell^2).
\end{eqnarray}

The single-particle Green's function is defined by
\begin{equation}
G^{nst}_{n's't'} = - \left< T \hat{c}_{nXst}(\tau)\hat{c}^\dagger_{n'X's't'}(0)\right>,
\end{equation}
and its Fourier-transform by
\begin{equation}
G^{nst}_{n's't'}(\bm{G},\tau)=\frac{2\pi\ell^2}{S}\sum_X G^{nst}_{n's't'}(X,X-G_y\ell^2,\tau)\exp(-iG_x X+\frac{1}{2}G_xG_y\ell^2).
\end{equation}
Within the HF approximation,
the equation of motion (EOM) for $G^{nst}_{n's't'}(\bm{G},\omega_m)$ is given by
\begin{equation}
(i\omega_m+\mu/\hbar) G^{nst}_{n's't'}(\bm{G},\omega_m)-
\frac{1}{\hbar}\sum_{t_1,s_1,n_3,\bm{G'}}
A_g(s,t,n,s_1,t_1,n_3;\bm{G},\bm{G'})G^{n_3s_1t_1}_{n's't'}(\bm{G'},\omega_m)
=\delta_{nn'}\delta_{ss'}\delta_{tt'}\delta_{\bm{G},0},
\label{eq:eom}
\end{equation}
where
\begin{eqnarray}
\label{Ag}
A_g(s,t,n,s_1,t_1,n_3;\bm{G},\bm{G'})&=&\varepsilon^s_n\delta_{n_3n}\delta_{t_1t}\delta_{s_1s}\delta_{\bm{G'}\bm{G}}\\ \nonumber
&&+[U_H(n,n_3;\bm{G'-G})\delta_{t_1t}\delta_{s_1s}-U_X(s_1,s,t_1,t,n,n_3;\bm{G'-G})]e^{i\bm{G}\times\bm{G'}\ell^2/2}.
\end{eqnarray}

\end{widetext}
Because LLM could be important, we retain several ``active'' LLs
(with LL indices between $n_{lower}$ and $n_{upper}$) around the
chemical potential (see Fig.~\ref{division}); i.e., we solve
the EOM explicitly for the Green's function matrix allowing
off-diagonal elements in the LL index for values $n$ satisfying
$n_{lower} \le n \le n_{upper}$.
However, it would be incorrect to completely neglect the filled
LLs below the active LLs ($n_c \leq n < n_{lower}$). These
levels can enter the calculations through $U_H$ and
$U_X$.  However if sufficiently below the chemical
potential, we expect LL mixing to be negligible for these states.
We thus treat these levels as ``inactive'', and
fix their density matrix elements to be
\(\left<\hat{\rho}^{nst}_{n's't'}(\bm{G})\right>=\delta_{nn'}\delta_{ss'}\delta_{tt'}\delta_{\bm{G},0}\).
(For self-consistency, we verify numerically that LL mixing for
the lowest active level is very small, justifying the dividing point
between active and inactive levels.)
With this form the inactive levels do not contribute to $U_H$ due to the
$(1-\delta_{\bm{G},0})$ in the Hartree term; i.e., it is precisely
cancelled by an interaction with a uniform neutralizing background.\cite{CoteR1991}
However, these
levels do contribute a non-vanishing exchange energy $U_X$,
\begin{eqnarray*}
\lefteqn{U_X^{inact}(s_1,s_2,t_1,t_2,n_2,n_3;\bm{G})=}\\
&&\frac{e^2}{4\ell}\Sigma_X(n_2,n_3)\delta_{s_1s_2}\delta_{t_1t_2}\delta_{\bm{G},0},
\end{eqnarray*}
where $inact$ stands for ``inactive'' and
\begin{equation}
\Sigma_X(n_2,n_3)=\sum_{n_1=n_c}^{n_{lower}-1}X_g(n_1,n_2,n_3,n_1;0).
\end{equation}
We can rewrite Eq.~(\ref{Ag}) as
\begin{eqnarray*}
\lefteqn{A_g(s,t,n,s_1,t_1,n_3;\bm{G},\bm{G'})}\\
&=&[\varepsilon^s_n\delta_{n_3n}-\frac{e^2}{4\ell}\Sigma_X(n,n_3)]
\delta_{t_1t}\delta_{s_1s}\delta_{\bm{G'}\bm{G}}\\ \nonumber
&&+[U_H^{act}(n,n_3;\bm{G'-G})\delta_{t_1t}\delta_{s_1s}\\
\nonumber
&&-U_X^{act}(s_1,s,t_1,t,n,n_3;\bm{G'-G})]e^{i\bm{G}\times\bm{G'}\ell^2/2},
\end{eqnarray*}
where the superscripts $act$ means now the summations in $U_H$ and
$U_X$  are restricted to the active LLs.

\begin{psfrags}
\psfrag{m}{$\mu$}
\psfrag{E}{$E$}
\psfrag{act}{active}
\psfrag{ina}{inactive}
\psfrag{ngltd}{neglected}
\psfrag{nc}{$n=n_c$}
\psfrag{nn1}{$n=-1$}
\psfrag{n0}{$n=0$}
\psfrag{n1}{$n=1$}
\psfrag{n2}{$n=2$}
\psfrag{n3}{$n=3$}
\psfrag{n4}{$n=4$}
\psfrag{n5}{$n=5$}
\psfrag{6}{$n=6$}
\psfrag{7}{$n=7$}
\begin{figure}
\includegraphics[scale=1.0]{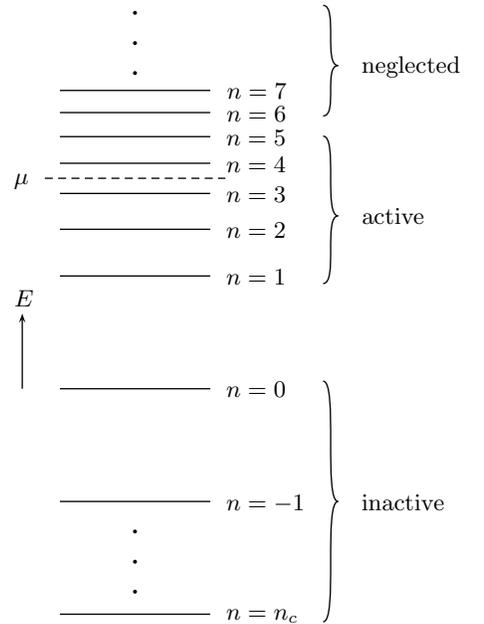}%
\caption{\label{division}Division of LLs into active and inactive LLs. In this example,
$n_{lower}=1$ and $n_{upper}=5$. The upper and lower cutoffs of the active LLs are
determined by the self-consistency conditions
\(1-\left<\hat{\rho}^{n_{lower}st}_{n_{lower}st}(0)\right>\,\ll 1\) and
\(\left<\hat{\rho}^{n_{upper}st}_{n_{upper}st}(0)\right>\,\ll 1\). Chemical potential $\mu$ for the case
of filling $\nu=14$ is indicated as
an example.  In general
several active levels are retained both above and below $\mu$
for all calculations reported here.
}
\end{figure}
\end{psfrags}

The calculations involve solving Eq.~(\ref{eq:eom}) to obtain
the Green's function, from which we obtain the density operator matrix elements.
Finally, the Hartree-Fock energy is given by
\begin{eqnarray*}
\lefteqn{E_{HF}}\\
&=&\frac{S}{2\pi\ell^2}\sum_{{n_2,n_3}\atop{(active)}}
\sum_{s,t}\lbrace[\varepsilon_{n_2}^s\delta_{n_2n_3}
-\frac{e^2}{4\ell}\Sigma_X(n_2,n_3)]\left<\hat{\rho}^{n_2st}_{n_3st}(0)\right>
\\ \nonumber &
&\quad +\frac{1}{2}\sum_{\bm{G}}[U_H^{act}(n_2,n_3;\bm{G})\left<\hat{\rho}^{n_2st}_{n_3st}(\bm{G})\right>
\\ \nonumber &
&\quad -\sum_{s_1,t_1}U_X^{act}(s_1,s_2,t_1,t_2,n_2,n_3;\bm{G})
\left<\hat{\rho}^{n_2st}_{n_3s_1t_1}(\bm{G})\right>]\rbrace \\
& &+\frac{S}{\pi\ell^2}\sum_{n\atop{(inact)}}[\sum_s \varepsilon_n^s-\frac{e^2}{4\ell}\Sigma_X(n,n)],
\end{eqnarray*}
where the last line is a constant for given $B$ and $n_{lower}$.

\section{\label{sec:result}results of the continuum limit model}
Table~\ref{occupation} details some typical results for
the occupations of the various Landau levels near the Fermi energy. In this example $\nu=14.5$;
i.e., the LL with
$(nst)=(4\uparrow\Uparrow)$ is half-filled. Note that in this notation
we denote the valley index as a pseudospin, with two values $\Uparrow$ and
$\Downarrow$ denoting the ${\bf K}$ and ${\bf K^{\prime}}$ valleys
respectively.
We present results for different coupling constants in the
range $0.5 \lesssim g \le 1$, consistent with previous
estimates of its appropriate value \cite{Dahal2007,Iyengar2007}.
Our qualitative results are very similar for different values of
$g$, even for (unphysical) values well above 1.
We can see that the occupations immediately
become very small above the half-filled LL, and very close to 1
below it, indicating that LLM is indeed small. This small level
of mixing, in spite of the small non-interacting energy gap between LLs where
the Fermi energy is located,  may be understood as being a consequence
of the large exchange enhancement of the gap due to the filled LLs.
Furthermore, for smaller $B$, deviations of the occupations
from a step function decreases (albeit just slightly), which means for decreasing field
LLM becomes even \textit{less} important. This
unintuitive result occurs because of the large sea of
negative energy LL states.  With
smaller field the degeneracy of each of these decreases, and so that more
inert LLs need to be filled to obtain the correct density
of electrons [see Eq.~(\ref{cutoff})]. In units of $e^2/\epsilon\ell$,
the exchange
interaction increases with increasing $|n_c|$, and the LLs effectively become
slightly more separated.
\begin{table}
\caption{\label{occupation}The diagonal density matrix  elements
\(\left<\hat{\rho}^{nst}_{nst}(0)\right>\), indicating the
occupation of the spin and pseudospin split LL with quantum numbers $(nst)$.
The occupations are very close to a step function,
indicating that LLM is weak. The deviation from a step
function decreases as $B$ decreases, indicating that LLM becomes
less important. Here $g$ is set to 1.}
\begin{ruledtabular}
\begin{tabular}{c c l l l}
\(n\)& \(st\) & \multicolumn{3}{c}{\(\left<\hat{\rho}^{nst}_{nst}(0)\right>\)}\\
\hline
     &        & \(n_c=1872\) &\(n_c=12000\) & \(n_c=24000\)\\
     &        & (\(B=20T\))   &(\(B=3.12T\)) & (\(B=1.56T\))\\
\hline
\(5\)&\((\downarrow\Downarrow),(\downarrow\Uparrow),(\uparrow\Downarrow)\)&\(0.0000697757\)&\(0.000059254\)&\(0.0000559402\)\\
\(5\)&\((\uparrow\Uparrow)\)                                               &\(0.000520925\)&\(0.000425574\)&\(0.000396389\)\\
\(4\)&\((\downarrow\Downarrow),(\downarrow\Uparrow),(\uparrow\Downarrow)\)&\(0.00078437\)&\(0.000651856\)&\(0.000611152\)\\
\(4\)&\((\uparrow\Uparrow)\)&\(0.499997\)&\(0.50002\)&\(0.500026\)\\
\(3\)&\((\downarrow\Downarrow),(\downarrow\Uparrow),(\uparrow\Downarrow)\)&\(0.999181\)&\(0.999318\)&\(0.999361\)\\
\(3\)&\((\uparrow\Uparrow)\)&\(0.999567\)&\(0.999627\)&\(0.999647\)\\
\end{tabular}
\end{ruledtabular}
\end{table}

\begin{figure}
\includegraphics[scale=0.75]{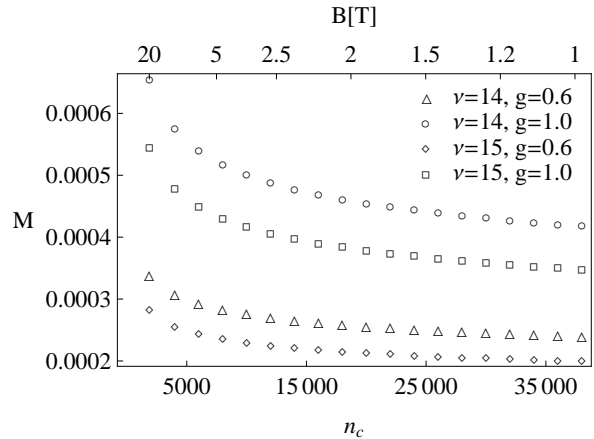}
\caption{\label{LLM}LLM for integer fillings.  Note $n_c \propto 1/B$, indicating
the LLM decreases with decreasing field.}
\end{figure}

Fig.~\ref{LLM} illustrates the LLM for two integer fillings where
the system is in a uniform liquid
state.  Here we measure the LLM by the quantity
\[M=\sum_{(nst)\neq (n's't')}\left<\hat{\rho}^{nst}_{n's't'}\right>^2,\]
where the sum is over active LLs only ($n_{lower}=-5$ and $n_{upper}=5$).
We again see that LLM is small and decreases as $B$ decreases for fixed filling factor.

Previous studies of crystal and stripe states in graphene in which a
single \cite{Dahal2007,ZhangCH2007,WangHao2007}
or small number \cite{ZhangCH2008} of Landau levels is retained
find that such states can be stable in the presence of
a magnetic field.  Our study suggests that inclusion
of the large number of LLs intrinsic to graphene not
only does not change such results, but even increases their validity
in weak fields.  The result of this is that, within a zero-temperature mean-field
description, one expects that in a very clean system many
different states will
persist down to the origin in a phase diagram plotted in the $n$ vs. $B$ plane
[see Fig.~\ref{fig:graphene})].  The state is determined only
by the filling factor.  This is in sharp contrast to the
situation for conventional 2DEG's, where LLM always destabilizes
such states as the origin is approached.

\begin{figure}
\begin{center}
\subfigure[]{\label{fig:wavelength}\includegraphics[scale=0.75]{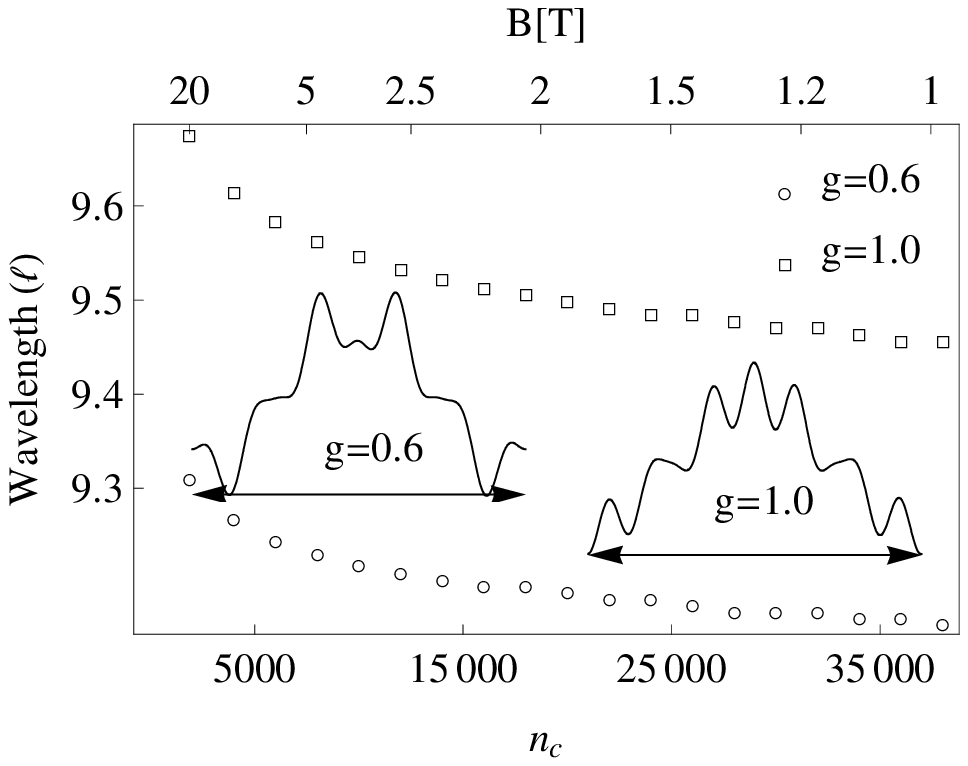}}
\subfigure[]{\label{fig:amplitude}\includegraphics[scale=0.75]{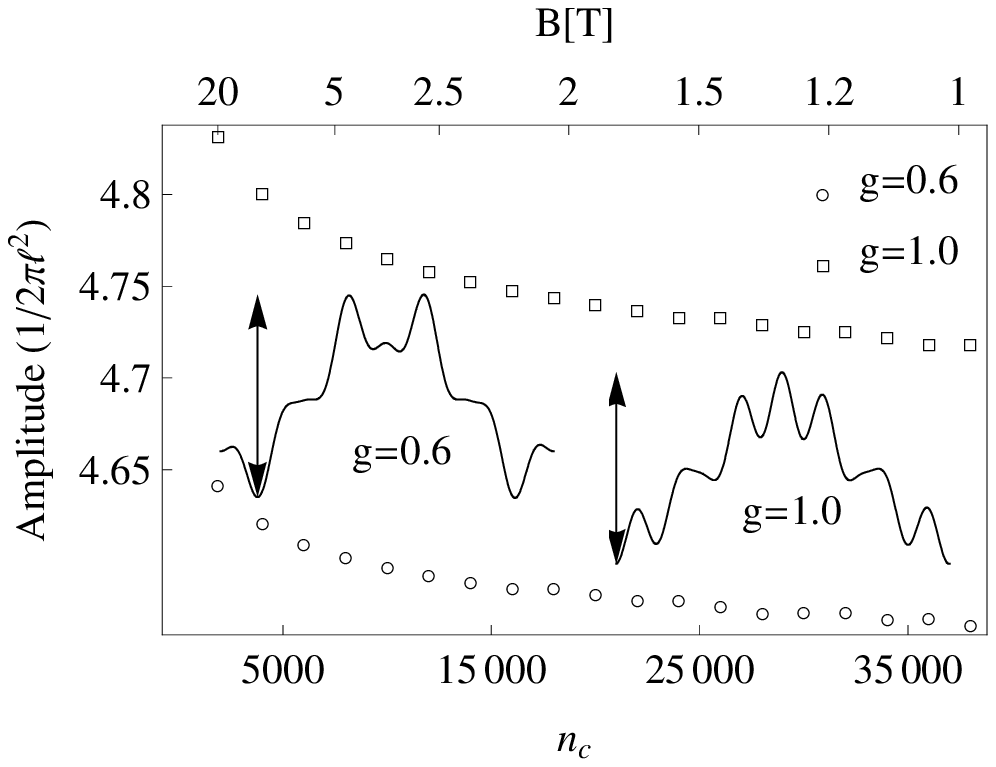}}
\end{center}
\caption{Wavelength and amplitude for stripe with
$\nu=14.5$,  $n_{lower}=1$ and $n_{upper}=5$, as functions of the
cutoff $n_c$ (and alternatively of the field $B$). The insets are
density profiles for different $g$'s, with the arrows indicating
the quantities plotted. } \label{fig:scaling}
\end{figure}

One seeming paradox associated with this behavior is
how the system approaches the uniform state which is believed,
at least within a mean-field approach,
to occupy the origin of the $n$ vs. $B$ phase diagram for graphene.
The answer lies in noting that since LL mixing is negligible, the
relevant length scale for the charge-ordered states of
a partially filled LL is the magnetic length, which diverges
as $B \rightarrow 0$.
Fig.~\ref{fig:scaling} illustrates the consequence of this for stripe
states.  One sees that the
wavelength and amplitude of the density modulation are basically
constants when measured in appropriate units ($\ell$ and
$1/2\pi\ell^2$, respectively).  Thus, these quantities should, up to logarithmic corrections, follow simple
scaling relations,
\begin{eqnarray}
\text{wavelength} \propto & \ell             &\propto  \frac{1}{\sqrt{B}},\\
\text{amplitude}  \propto & 1/\ell^2 &\propto  B.
\end{eqnarray}
As $B$ decreases, the stripes, and we believe CDWs in general, ``evaporate'',
and thus approach the expected uniform density state at the origin.

\begin{psfrags}
\psfrag{1}[cc]{1}
\psfrag{2}[cc]{2}
\psfrag{3}[cc]{3}
\psfrag{4}[cl]{4}
\psfrag{5}[cc]{5}
\psfrag{6}[cc]{6}
\psfrag{7}[cc]{7}
\psfrag{8}[cl]{8}
\psfrag{10}[cl]{$1'$}
\psfrag{20}[cl]{$2'$}
\psfrag{30}[cl]{$3'$}
\psfrag{40}[cl]{$4'$}
\psfrag{50}[cl]{$5'$}
\psfrag{60}[cl]{$6'$}
\psfrag{70}[cl]{$7'$}
\psfrag{80}[cl]{$8'$}
\psfrag{x}{x}
\psfrag{y}{y}
\begin{figure}
\includegraphics[scale=0.65]{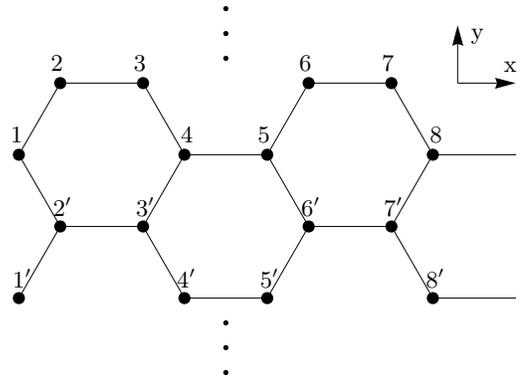}%
\caption{\label{muc}Geometry of the tight binding problem. The unit cell corresponds to the area between the two armchair chains, including one of the bounding chains. In this example the unit cell contains 8 sites and 4 plaquettes. In our calculations, there are $n_y$ unit cells in the y direction, but only one in the x direction. We apply periodic boundary conditions to both x and y directions, so that site 8 is connected to site 1 by the bond to its right, etc..
}
\end{figure}
\end{psfrags}

\section{\label{sec:hubbard}hubbard model}
The continuum limit forces one to adopt a
cutoff in the occupied states, which in the previous
section was accomplished by adopting an appropriate
choice of the minimum occupied LL index, $n_c$.
Since the increase of $n_c$ with decreasing
field tends to suppress LLM, one may wish to
consider whether a more physical cutoff scheme
would give similar results.
Towards this end we re-examine this question within a tight-binding
Hubbard model. As in the continuum case, we look for
states of this system within
the Hartree-Fock approximation. For a simple on-site interaction $U$,
the HF Hamiltonian
for
spin up electrons is
\begin{equation}
\hat{H}_{HF}(\uparrow)=\sum_{<ij>}t_{ij}\hat{a}^\dagger_{i\uparrow}\hat{a}_{j\uparrow}+
U\sum_{i}\left<\hat{n}_{i\downarrow}\right>\hat{a}_{i\uparrow}^\dagger
\hat{a}_{i\uparrow},
\label{eq:hubb_hf}
\end{equation}
where $<ij>$ indicates nearest neighbors.
For spin down electrons the Hamiltonian is analogous,
with $\uparrow$ and $\downarrow$ interchanged.

We choose the unit cell to be the area between two adjacent armchair chains (see Fig.~\ref{muc}). We apply periodic boundary conditions to both x and y directions and study stripes oriented along the y directions. We can Fourier transform along the y direction, then we only need to define the phases of $t_{ij}$ along one armchair chain (e.g., the chain $12345678$ in Fig.~\ref{muc}), i.e., $t_{ij}\rightarrow t_{i,i+1}$, where $i=1,2,\ldots, n_x$ labels the sites in the unit cell ($n_x=8$ for the example in Fig.~\ref{muc} and in general can be any integer multiple of 4). One possible choice is
\[ \arg t_{i,i+1}=\left\{
\begin{array}{l l}
  0 & \quad \mbox{if $i$ is even}\\
  (-1)^{(i-1)/2}(i-1)\pi\alpha & \quad \mbox{if $i$ is odd}\\ \end{array} \right. , \]
where $\alpha=2\Phi_m/n_x$ with $\Phi_m$ being the total number of flux quanta in the unit cell. (See Table~\ref{phases}.) Since $\Phi_m$ must be an integer, the magnetic fields for computationally tractable system sizes are actually very large compared to experiments. Nevertheless, we can deduce the qualitative behavior from these calculations.

\begin{table}
\caption{\label{phases} Nonzero $\arg t_{i,i+1}$ for hopping from site $i$ to $i+1$ on the chain $12345678$ in Fig.~\ref{muc}.}
\begin{ruledtabular}
\begin{tabular}{c c c c }
i & 3 & 5 & 7\\
\hline
$\arg t_{i,i+1}$ & $-2\pi\alpha$ & $4\pi\alpha$ & $-6\pi\alpha$\\
\end{tabular}
\end{ruledtabular}
\end{table}

The coupling constant in this
model is $g=U\ell/ta$, where  $t\approx2.7\,\text{eV}$ is the
hopping amplitude of the tight-binding approximation. The
situation that $g$
is field-independent does not arise naturally here; we introduce it
by adjusting $U$ with field
according to the relation
\[U\propto 1/\ell \propto \sqrt{B} \propto \sqrt{\frac{\Phi_m}{n_x}}.\]

For real Coulomb interactions, the effective HF potential
includes a short-range exchange potential and
a long-range Hartree potential, both proportional
to $\sqrt{B}$.  Stripe and bubble
states result from the competition of these \cite{koulakov06,moessner06}.
Because of the highly local nature of the interaction in the Hubbard
model, neither this scaling nor the effective long-range
part of the interaction emerge: one only finds a local repulsion
between electrons of different spins.  Thus, the charge-ordered
bubble and stripe states are not eigenstates of Eq.~(\ref{eq:hubb_hf}):
one generically finds uniform density states.  To obtain the former states,
one needs to include longer-range interactions in the Hamiltonian.
Obtaining full solutions of the HF approximation in this situation
is possible but challenging, and is unnecessary for our more modest
goal of testing the effect of using a real lattice rather than
an energy cutoff.  Thus, rather than fully solving for states of a system
with long-range interactions,
we include a slowly varying
external potential which models the effect of the
long range (non-contact) part of the potential.
For simplicity we take this to have the form
\[\Delta \hat{H}(\uparrow)=A \sum_{i} \cos(2\pi x_i/L_x) \hat{a}_{i\uparrow}^\dagger \hat{a}_{i\uparrow},\]
where $A$ must scale with field in the same way as $U$, and $L_x$ is the length of the unit cell along the x direction.

\begin{figure}[t]
\includegraphics[scale=0.75]{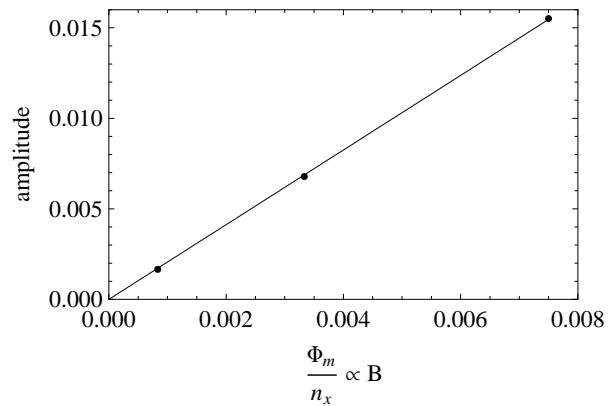}
\caption{\label{fig:tb}  Stripe amplitude for Hubbard model
calculation for a fixed ratio of unit cell width to
magnetic length, as a
function of $\Phi_m/n_x$, which is proportional to the field. 
$U/t=76.2\sqrt{\Phi_m/n_x}$, $A/U=0.1$.
A
straight line through the data points extrapolates
rather accurately through the origin.}
\end{figure}

Our goal is to study how the density of a CDW state varies if the field
is allowed to change, keeping the effective $g$ fixed.
In order to make a fair comparison between states at different
field strengths,
we also fix the ratio \(\frac{n_x}{\ell}\propto \sqrt{n_x
\Phi_m}\) so that the width of the stripes and their
spacing relative to the unit cell size does not change.
This restricts the number of systems we can examine. However, the
data we do get are in excellent agreement with the continuum
model, i.e., the stripe amplitude (defined as the difference in maximum and minimum densities)
is roughly proportional to the magnetic field (see Table~\ref{tb} and Fig.~\ref{fig:tb}).
Note that in these calculations the amplitude decreases
slightly faster than linearly with the field, consistent
with a decreasing role for Landau level mixing.  This effect 
is larger for larger values of $A/U$, as illustrated for example
in Table~\ref{tb}.

\begin{table}
\caption{\label{tb} Change in stripe amplitude,
defined as the difference between the maximum and minimum densities
within a unit cell,
when the magnetic field is changed by varying $\Phi_m$ and $n_x$.
For first row of data, $A/U=0.2$; for second row, $A/U=0.15$.
For all data, $U/t=76.2\sqrt{\Phi_m/n_x}$.
}
\begin{ruledtabular}
\begin{tabular}{c c c c }
\( (\Phi_m,\,n_x)\)& \( (\Phi_m',\,n_x')\) & \(B'/B\) & \(\text{amplitude}'/\text{amplitude}\) \\
\hline
(1~,~600) & (2~,~300) & 4 & 4.1941\\
(1~,~800) & (2~,~400) & 4 & 4.1509\\
\end{tabular}
\end{ruledtabular}
\end{table}

In Fig.~\ref{fig:tb} we illustrate the stripe amplitude for states
generated for three values of \((\Phi_m,\,n_x)\), corresponding to
three different magnetic fields, but with the ratios of the unit
cell sizes and magnetic length the same, and with a relatively
small value of $A/U$ (0.1).  In this case
one may fit a straight line
through these points, and find that it extrapolates to the origin
rather accurately.  This is consistent with the stripe amplitude
continuously vanishing in the $B \rightarrow 0$ limit, as was
found in the continuum approach.

\section{\label{sec:sum}summary}
We have examined the stability of liquid and charge-ordered
states for graphene (focusing
on stripes as a paradigm for the latter)
in the quantum Hall regime against the effects
of Landau level mixing.  Because the
coupling constant $g$ is independent of field, we find
the LLM does not increase with decreasing field,
and that, counterintuitively, it decreases, albeit by
a small amount.
This latter effect
is due to
a large exchange enhancement of the LL gaps from the filled
negative energy LLs, which increase in number as the field
decreases.  Within mean-field theory, this implies that
clean and cold graphene at small fields and densities
should support many different phases,
determined solely by the filling factor.
This contrasts with the conventional 2DEG, where
a Wigner crystal state is believed to reside
throughout this regime.  In graphene, the liquid phase
thought to exist in the absence of doping is reached
in the $B \rightarrow 0$ limit at fixed filling
factor by an ``evaporation'' of the CDW, in which
the amplitude vanishes linearly with $B$.

\begin{acknowledgments}
The authors thank M. Fogler, R. C\^ot\'e and I. Herbut for helpful
discussions. This work is supported by NSF Grant No. DMR-0704033 and MAT2006-03741(Spain).
Computer time was provided by Indiana University.
\end{acknowledgments}


\begin{thebibliography}{27}
\bibitem[{com()}]{com}
\bibinfo{note}{We assume that the electrons in the negative levels can be
  described as filling an integral number of LLs. Although not precisely true,
  the discreteness of $n_c$ has little effect because in realistic situations
  $n_c$ is very large.}
\end{thebibliography}
\end{document}